\documentstyle[epsfig,mathptm]{mn}
\DeclareMathVersion{bold}
\newif\ifAMStwofonts
\AMStwofontstrue                
\def\textbfit{\protect\txtbfit}

\long\def\txtbfit#1{{\fontfamily{cmr}\fontseries{bx}\fontshape{it}%
  \selectfont #1}}

\DeclareMathVersion{bold}                    
\DeclareMathAlphabet{\mathbfit}{OT1}{cmr}{bx}{it}
\SetMathAlphabet\mathbfit{bold}{OT1}{cmr}{bx}{it}
\DeclareMathAlphabet{\mathbfss}{OT1}{cmss}{bx}{n}
\SetMathAlphabet\mathbfss{bold}{OT1}{cmss}{bx}{n}
\ifAMStwofonts
  \ifCUPmtlplainloaded \else
    \DeclareSymbolFont{UPM}{U}{eur}{m}{n}
    \SetSymbolFont{UPM}{bold}{U}{eur}{b}{n}
    \DeclareSymbolFont{AMSa}{U}{msa}{m}{n}
    \DeclareMathSymbol{\upi}{0}{UPM}{"19}
    \DeclareMathSymbol{\umu}{0}{UPM}{"16}
    \DeclareMathSymbol{\upartial}{0}{UPM}{"40}
    \DeclareMathSymbol{\leqslant}{3}{AMSa}{"36}
    \DeclareMathSymbol{\geqslant}{3}{AMSa}{"3E}

    \let\leq=\leqslant 
    \let\geq=\geqslant 
  \fi
\fi
\title[Wavelets and non-Gaussianity]
{Wavelet analysis and the detection of non-Gaussianity in the CMB}
\author[M.P. Hobson, A.W. Jones and A.N.~Lasenby]
{M.P. Hobson, A.W. Jones and A.N.~Lasenby\\
Astrophysics Group, Cavendish Laboratory, Madingley Road, 
Cambridge CB3 OHE, UK\\}
\date{Accepted ???. Received ???; in original form 10 October 1998}
\pagerange{\pageref{firstpage}--\pageref{lastpage}}
\pubyear{1998} 

\begin{document} 
\maketitle 
\label{firstpage}

\begin{abstract}
We investigate the use of wavelet transforms in detecting and
characterising non-Gaussian structure in maps of the cosmic microwave
background (CMB).  We apply the method to simulated maps of the Kaiser-Stebbins
effect due to cosmic strings onto which Gaussian signals of varying
amplitudes are superposed. We find the method significantly
outperforms standard techniques based on measuring the moments of the
pixel temperature distribution. We also compare the results with
those obtained using techniques based on Minkowski functionals, and we
again find the wavelet method to be superior. In particular, 
using the wavelet technique, we find that it
is possible to detect non-Gaussianity even in the presence of a
superposed Gaussian signal with five times the rms amplitude of the
original cosmic string map. We also find that the wavelet technique is
useful in characterising the angular scales at which the non-Gaussian
signal occurs.
\end{abstract}


\section{Introduction}
\label{intro}

The cosmic microwave background (CMB) is widely considered as one of 
the most important experimental tools in investigating the formation of
structure in the Universe. In particular, by observing the temperature
fluctuations in the CMB, we hope to distinguish between two competing
theoretical paradigms.  The first is the standard
inflationary model, incorporating cold dark matter, for which
the distribution of temperature fluctuations in the CMB should be
Gaussian.  The second class of theories 
invoke the formation of topological defects such as cosmic strings,
monopoles or textures. In these theories the CMB temperature
fluctuations are expected to be non-Gaussian, possessing steep
gradients or well-defined `hot spots' of emission (Bouchet, Bennett \&
Stebbins 1988; Turok 1996). Thus the detection (or otherwise) of a
non-Gaussian signal in the CMB is an important means of discriminating
between these two classes of theory.

Even in defect theories, however, it is possible that the non-Gaussian
CMB signal may be obscured by an additional Gaussian component
of primordial CMB fluctuations, by foreground emission from the
Galaxy, or by instrumental noise. Moreover, CMB
fluctuations arising from realistic defect cosmologies 
may well contain a Gaussian component that has 
a similar power spectrum to that of the non-Gaussian component
(Magueijo \& Lewin 1997),
thereby making the task of detecting the non-Gaussian signal
even more difficult. It is therefore important 
to investigate the regimes in which
underlying non-Gaussianity can be detected when additional
Gaussian signals are present. The most common technique is to define a
statistic with properties that are easily calculated for Gaussian
processes and then apply this statistic to the (possibly) non-Gaussian
CMB map. In this way it is often possible to determine the probability
that a given test map is drawn from an underlying Gaussian
ensemble. Several approaches to this problem have
already been proposed. Perhaps the most straightforward of these is the
measurement of the skewness and kurtosis of the distribution of pixel
temperatures in the map (Scaramella \& Vittorio 1991).  More elaborate
methods include the calculation of the 3-point correlation function of
the temperature distribution (Kogut et al. 1996) and the investigation
of the statistics of maxima and minima in the map (Bond \& Efstathiou
1987; Vittorio \& Juskiewicz 1987).  Methods based on the topological
properties of the two-dimensional CMB temperature distribution have
also been proposed (Coles 1988; Gott et al. 1990). The most recent of
these techniques is based on the calculation of the three Minkowski
functionals of the temperature distribution (see Section
\ref{minkfunc}).

In this paper, we investigate the use of a new method for detecting
non-Gaussianity which is based on the idea of expressing the CMB
temperature distribution in terms of a set of two-dimensional wavelet
basis functions (Fang \& Pando 1996; Ferreira, Magueijo \& Silk 1997).
The wavelet transform provides a natural decomposition of the image 
into structure on different scales and, by analysing the distribution of
wavelet coefficients on a given scale, it is possible to avoid the
usual restrictions imposed by the central limit theorem, so that 
non-Gaussian signal is more easily detected. 
We apply this technique to simulated maps of
the CMB that contain both a non-Gaussian contribution, due to the
Kaiser-Stebbins effect from cosmic strings (Kaiser \& Stebbins 1984),
and also a Gaussian component
with an identical power spectrum to the strings map. By varying the
relative amplitudes of the two contributions, we determine to what extent the
non-Gaussian signal can be obscured while still remaining detectable
by the wavelet method. We compare our results with those obtained
using either Minkowski functionals or straightforward calculation of
the moments of the pixel temperature distribution.

\section{The discrete wavelet transform}
\label{wavelets}

The discrete wavelet transform (DWT) has been discussed extensively
elsewhere (e.g. Daubechies 1992).  In particular, a characteristically
clear introduction is given by Press et al. (1994). We therefore give
only a brief description of its properties.

\subsection{The one-dimensional DWT}
\label{wave1d}

It is best initially to discuss the DWT in just one dimension.
In order to develop an intuitive understanding,
let us begin by considering a periodic
function $f(x)$ of period $L$ 
that we wish to expand in terms of a {\em wavelet basis}.  The construction
of the wavelet basis begins with the specification of 
the mother (or analysing) wavelet $\psi(x)$, together with another
related function $\phi(x)$ called the father wavelet (or
scaling function). In order to construct a wavelet basis that is
discrete, orthogonal and compactly-supported, 
the functions $\phi(x)$ and $\psi(x)$ must
together obey several highly-restrictive mathematical relations first derived
by Daubechies (1992). We will not list these relations here, but merely
note that the most straightforward requirements are that
\begin{eqnarray}
\int_{-\infty}^{\infty} \phi(x)\,{\rm d}x & = & 1,\label{phiint} \\
\int_{-\infty}^{\infty} \psi(x)\,{\rm d}x & = & 0.\label{psiint}
\end{eqnarray}
Mother and father functions with
compact support are usually defined on the
interval $[0,1]$, and the
wavelet basis is then constructed from dilations and translations
of $\phi(x)$ and $\psi(x)$ as
\begin{eqnarray}
\phi_{j,l}(x) & = & \left(\frac{2^j}{L}\right)^{1/2}\phi(2^jx/L-l),
\label{phibasis} \\
\psi_{j,l}(x) & = & \left(\frac{2^j}{L}\right)^{1/2}\psi(2^jx/L-l),
\label{psibasis}
\end{eqnarray}
where $j$ and $l$ are integers. The index $j$ labels the scale size
of the wavelet, whereas $l$ labels the position of the wavelet
at this scale. The set $\{\phi_{0,l},\psi_{j,l}\}$ with 
$0\leq j < \infty$ and $-\infty < l < \infty$ forms a complete, orthonormal
basis in the space of functions of period $L$. The orthogonality
of these functions is formally given by
\begin{eqnarray*}
\int_{-\infty}^{\infty} \phi_{j,l}(x)\phi_{j,l'}(x)\,{\rm d}x   & = &
\delta_{ll'},\\
\int_{-\infty}^{\infty} \phi_{j,l}(x)\psi_{j',l'}(x)\,{\rm d}x  & = &
\delta_{jj'}\delta_{ll'},\\
\int_{-\infty}^{\infty} \psi_{j,l}(x)\psi_{j',l'}(x)\,{\rm d}x  & = &
\delta_{jj'}\delta_{ll'}.
\end{eqnarray*}
Using these orthogonality relationships, we may then write the
function $f(x)$ as
\begin{equation}
f(x) = \sum_{l=-\infty}^{\infty} a_{0,l}\phi_{0,l}(x)
+ \sum_{j=0}^{\infty}\sum_{l=-\infty}^{\infty} b_{j,l}\psi_{j,l}(x),
\label{wtcont}
\end{equation}
where the {\em wavelet coefficients} $a_{0,l}$ and $b_{j,l}$ are given by
\begin{eqnarray}
a_{0,l} & = & \int_{-\infty}^{\infty} f(x) \phi_{0,l}(x)\,{\rm d}x,
\label{acoeffs}\\
b_{j,l} & = & \int_{-\infty}^{\infty} f(x) \psi_{j,l}(x)\,{\rm d}x.
\label{bcoeffs}
\end{eqnarray}

So far we have assumed our function $f(x)$ is defined for all values of $x$.
In many applications, however, the function is in fact digitised
(sampled) at $N=2^J$ equally-spaced
points $x_i$. In this case, 
by analogy with the Discrete Fourier Transform (DFT) of a sampled
function, (\ref{wtcont}) becomes
\begin{equation}
f(x_i) = a_{0,0}\phi_{0,0}(x_i)
+ \sum_{j=0}^{J-1}\sum_{l=0}^{2^j-1} b_{j,l}\psi_{j,l}(x_i),
\label{wtdig}
\end{equation}
and the integrals for the wavelets coefficents in (\ref{acoeffs}) \&
(\ref{bcoeffs}) are replaced by the appropriate summations. If the
mean of the function samples, $f(x_i)$,
is zero (as will be case in this paper), then
$a_{0,0}=0$ and the function can be described entirely in terms of the
wavelets $\psi_{j,l}$. As the scale index $j$ increases from 0 to
$J-1$, the wavelets represent the structure of the function on
increasingly smaller scales, with each scale a factor of 2 finer than
the previous one.  The index $l$ (which runs from 0 to $2^j-1$)
denotes the position of the wavelet $\psi_{j,l}$ within the $j$th
scale level.

If we collect the function samples $f_i=f(x_i)$ in a column vector
${\mathbfss f}$ (whose length $N$ must an integer power of 2), then
(like the DFT) the DWT is a linear operation that transforms 
${\mathbfss f}$ into another vector $\widetilde{\mathbfss f}$ of
the same length, which contains the wavelet coefficients of 
the (digitised) function. The action of the DWT can
therefore be considered as the multiplication of the original
vector by the $N\times N$ wavelet matrix ${\mathbfss W}$ to give
\begin{equation}
\widetilde{\mathbfss f} = {\mathbfss Wf}.
\label{dwt1d}
\end{equation}
Again like the DFT, the wavelet matrix ${\mathbfss W}$ is
orthogonal, so the inverse transformation can be performed
straightforwardly using the transpose of ${\mathbfss W}$.  Thus both
the DFT and DWT can be considered as rotations from the original
orthonormal basis vectors ${\mathbfss e}_i$ in signal space to some new
orthonormal basis $\widetilde{\mathbfss e}_i$ ($i=1,\ldots,N$), with the
transformed vector $\widetilde{\mathbfss f}$ containing the coefficients in
this new basis.

The original basis vectors ${\mathbfss e}_i$ have unity as the $i$th
element and the remaining elements equal to zero, and hence correspond
to the `pixels' in the original vector ${\mathbfss f}$. Therefore
the original basis is the most localised basis possible in real
space. For the DFT, the new basis vectors $\widetilde{\mathbfss e}_i$ are
(digitised) complex exponentials and represent the opposite extreme,
since they are completely non-local in real space but localised in
frequency space. For the DWT, the new basis vectors are the
wavelets, which enjoy the characteristic property of being reasonably
localised both in real space and in frequency space, thus occupying an
intermediate position between the original `delta-function' basis
and the Fourier basis of complex exponentials. Indeed, it is
the simultaneous localisation of wavelets in both spaces that makes 
the DWT such a useful tool for analysing data in wide range of applications.

Since the wavelet basis $\widetilde{\mathbfss e}_i$ consists of (digitised)
dilations and translations of the mother and father wavelets 
$\psi(x)$ and $\phi(x)$, we can obtain a different wavelet basis for each
pair of these functions that obey the Daubechies conditions.
Thus there exists an infinite number of possible wavelet transforms,
with different wavelet bases making different trade-offs between how
compactly they are localised in either real space or frequency space. In
this paper, we will be concerned only with wavelet bases for
which the wavelets are continuous and have compact support in either
space (it is impossible for a function to have
compact support in both spaces).

The implementation of the DWT is based on a pyramidal algortihm
(Daubechies 1992; Press et al. 1994), and its speed is comparable to
a Fast Fourier Transform. The calculation is arranged so that the transformed
signal vector $\widetilde{\mathbfss f} = {\mathbfss Wf}$, containing
the wavelet coefficients, takes the form
\[
\widetilde{\mathbfss f} = (a_{0,0},b_{0,0},b_{1,0},b_{1,1},
b_{2,0},b_{2,1},b_{2,2},b_{2,3},\ldots)^{\rm t}.
\]
Thus the wavelet coefficients are placed in order of increasing
scale index $j$ and, within a given scale, the $2^j$ coefficients are
ordered with increasing position index $l$.

\subsection{The two-dimensional DWT}
\label{wave2d}

The extension of the DWT to two-dimensional signals (or images) is
straightforward and is usually performed by
taking simple tensor products of the one-dimensional wavelet basis
to produce a two-dimensional wavelet basis given by
\begin{eqnarray*}
\phi_{0,0;l_1,l_2}(x,y) & = & \phi_{0,l_1}(x)\phi_{0,l_2}(y) \\
\zeta_{j_1,0;l_1,l_2}(x,y) & = & \psi_{j_1,l_1}(x)\phi_{0,l_2}(y) \\
\xi_{0,j_2;l_1,l_2}(x,y) & = & \phi_{0,l_1}(x)\psi_{j_2,l_2}(y) \\
\psi_{j_1,j_2;l_1,l_2}(x,y) & = & \psi_{j_1,l_1}(x)\psi_{j_2,l_2}(y).
\end{eqnarray*}
If the two-dimensional pixelised image has dimensions $2^{J_1}\times
2^{J_2}$ then by analogy with (\ref{wtdig}) we have 
$0 \leq j_1 \leq J_1-1$ and $0 \leq l_1 \leq 2^{j_1}-1$, and similarly
for $j_2$ and $l_2$.

\begin{figure}
\centerline{\epsfig{
file=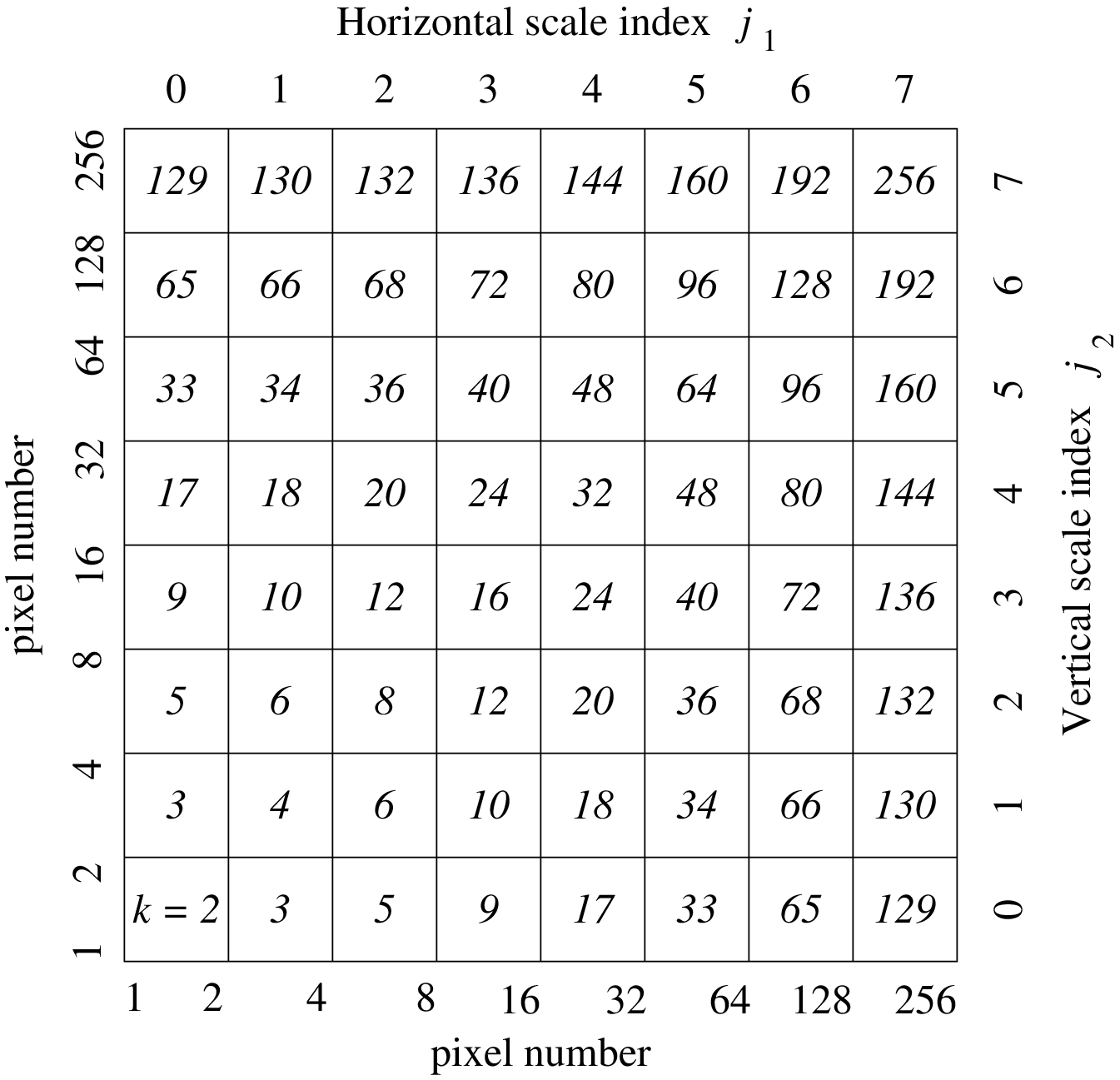,width=8cm}}
\caption{The domains of the matrix $\widetilde{\mathbfss T}$
containing the coefficients of the two-dimensional wavelet transform.
The italic numbers show the value of
$k=2^{j_1}+2^{j_2}$ in each domain; see text for details.}
\label{domains}
\end{figure}

In this paper, will we be concerned with two-dimensional 
pixelised maps of temperature fluctuations in the 
CMB with dimensions $256 \times 256$ (i.e. $J_1=J_2=8$).
If we denote such an image by the matrix ${\mathbfss T}$, then the 
matrix of wavelet coefficients is given by
\begin{equation}
\widetilde{\mathbfss T} = {\mathbfss WTW}^{\rm t},
\label{dwt2d}
\end{equation}
where ${\mathbfss W}$ is the wavelet matrix introduced in (\ref{dwt1d}) and
${\mathbfss W}^{\rm t}$ is its transpose. 
The structure of the
matrix $\widetilde{\mathbfss T}$ containing the wavelet coefficients
is shown in Fig.~\ref{domains},
where the pixel numbers are plotted on a logarithmic scale.
We see that the matrix is partitioned into separate domains according
to the scale indices $j_1$ and $j_2$ in the horizontal and vertical
directions respectively. By analogy with the one-dimensional case,
as $j_1$ increases the wavelets represent the horizontal structure in
the image on increasingly smaller scales. Similarly, as $j_2$ increases 
the wavelets represent the increasingly fine scale vertical structure in the
image. Thus domains that lie in the leading diagonal in Fig.\ref{domains}
(i.e. with $j_1=j_2$) contain coefficients of two-dimensional 
wavelets that represent the image at the same scale in the horizontal
and vertical directions, whereas domains with $j_1 \neq j_2$
contain coefficients of two-dimensional wavelets describing the image on
different scales in the two directions.

Finally, we note that for the two-dimensional DFT it is common to denote
the distance from the origin in Fourier space by $k$, 
which serves as a measure of inverse scale-length. 
In a two-dimensional DWT, it is also useful to define a
similar quantity, but in this case $k$ is an integer variable given by
$k=2^{j_1}+2^{j_2}$ which takes certain prescribed values between
$k_{\rm min} = 2$ and $k_{\rm max}=2^{J_1-1}+2^{J_2-1}$. In our
application $J_1=J_2=8$ and so $k_{\rm max}=256$.  By definition, the
value of $k$ is constant within each of the domains shown in
Fig.\ref{domains}.  The values of $k$ in each domain are shown as the
italic numbers in Fig.~\ref{domains}.  We note that $k$ takes the same
value for domains lying symmetrically on either side of the leading
diagonal.

\section{Detecting non-Gaussianity with the DWT}
\label{wtngdec}

As mentioned above, let us denote our two-dimensional pixelised test
map of temperature fluctuations in the CMB by the matrix ${\mathbfss
T}$. If the test map is not drawn from a Gaussian ensemble, then we might
hope to measure this non-Gaussianity directly from the
histogram of the pixel temperatures, i.e. the histogram of the
elements of the matrix ${\mathbfss T}$. Indeed, Scaramella \& Vittorio
(1991) use the skewness and kurtosis of the pixel temperature
distribution as statistics for detecting non-Gaussianity in CMB images.

As noted by Fang \& Pando (1996), however, methods based on statistics of the
pixel temperature distribution can be unreliable in detecting
non-Gaussianity, as a result of the central limit theorem.  For
example, if an image consists of numerous non-Gaussian processes on
different scale lengths, then from the central limit theorem we would
expect the probability distribution of the pixel temperatures to tend
to a Gaussian as the number of processes increases.

In this paper, therefore, rather than analysing the distribution of
the elements of ${\mathbfss T}$, we concentrate instead on the
statistics of the wavelet coefficients, contained in the matrix
$\widetilde{\mathbfss T}$.  
The wavelet transform
provides a natural decomposition of the image into structure on
different scales, and within each scale the wavelet basis functions
are well-localised. Thus, by analysing the distribution of
wavelet coefficients on a given scale, it is possible to avoid the
restriction of the central limit theorem and any non-Gaussian signal
on this scale is far more pronounced.

\subsection{Statistics of wavelet coefficients}
\label{wtstats}

Referring to Fig.~\ref{domains}, it is
useful to consider together all the wavelet coefficients that lie in
domains sharing the same value of $k$. It is straightforward to show (and is
intuitively reasonable) that if the input image ${\mathbfss T}$ is
(cyclically) translated or rotated, then wavelet coefficients sharing
the same $k$-value are merely rearranged within the relevant domains.
Thus any statistic based on the distribution of all wavelet
coefficients sharing the same $k$-value is invariant to translations
and rotations of the input image.

In order to describe the statisics of the wavelet coefficients of a
given test image, we could (for each value of $k$
separately) calculate estimators of the moments $\hat{\nu}_r$
($r=1,2,3,\ldots$), or central moments $\hat{\mu}_r$, of the 
distributions of its wavelet
coefficients (see the Appendix).  In this way, we
obtain {\em moment spectra} $\hat{\nu}_r(k)$ (or central
moment spectra $\hat{\mu}_r(k)$) for $r=1,2,3,\ldots$ which describe
the (scale dependent) statistics of the wavelet coefficients.  As
discussed by Ferreira et al. (1997), however, a better approach is to
describe the distribution of wavelet coefficients at each $k$-value in
terms of its cumulants.  Using unbiassed estimators of the cumulants
based on $k$-statistics (see the Appendix), we thus calculate
{\em cumulant spectra} $\hat{\kappa}_r(k)$ ($r=1,2,3,\ldots$)
of the wavelet coefficients of the test image.

For a given test image ${\mathbfss T}$, the cumulant spectra of the
corresponding wavelet coefficients have some useful
properties. Firstly, let us suppose that ${\mathbfss T}$ is drawn from
a Gaussian ensemble. Since the wavelet transform can be considered
merely an orthogonal rotation of the basis vectors in image space, it
is straightforward to show that the wavelet
coefficients at each value of $k$ are also drawn from a
Gaussian distribution. Thus, for Gaussian
test images, the expectation value of the estimated cumulant spectra
$\hat{\kappa}_r(k)$ of the wavelet coefficients is zero for $r > 2$
(see the Appendix). Secondly, let us consider a test image 
${\mathbfss T}={\mathbfss T}_1+{\mathbfss T}_2$ that is the
sum of two independent processes ${\mathbfss T}_1$ and ${\mathbfss T}_1$.
Since the DWT is a linear operation, we
thus have $\widetilde{\mathbfss T} =\widetilde{\mathbfss
T}_1+\widetilde{\mathbfss T}_2$. Using the additive property of
cumulants discussed in the Appendix, we therefore find that the $r$th cumulant
spectrum $\hat{\kappa}_r(k)$ of the wavelet coefficients of
${\mathbfss T}$
is simply the sum of the corresponding cumulant spectra of the wavelet
coefficients of ${\mathbfss T}_1$ and ${\mathbfss T}_2$ taken separately.
Of course, this argument can be extended to an arbitrary number of
independent processes.

The cumulant spectra $\hat{\kappa}_r(k)$ ($r=1,2,3,\ldots$) form our
basic test for non-Gaussianity. The procedure is as follows. We
begin with some (possibly non-Gaussian) test image ${\mathbfss T}$ and
perform a wavelet transform to obtain the corresponding matrix of
wavelet coefficients $\widetilde{\mathbfss T}$ given by (\ref{dwt2d}). We
then calculate the cumulant spectra $\hat{\kappa}_r(k)$ of
these wavelet coefficients.  Returning to the original test image
${\mathbfss T}$, we then create a large number (5000 say) of {\em
equivalent Gaussian realisations} (EGR).  Each EGR is created by Fourier
transforming the original test image, randomising the phases of the
complex Fourier coefficients, and inverse Fourier transforming the
result. In the randomisation step, the phase of each Fourier
coefficient is drawn at random from a uniform distribution in
the range $[0,2\pi]$. Thus each EGR has exactly the same power spectrum
as the original test image, but the pixel temperature distribution is
drawn from a Gaussian ensemble. Each EGR is then analysed in exactly
the same way as the original test image in order to obtain the
cumulant spectra of its wavelet coefficients. Thus, after analysing
all the EGRs, we obtain (for each value of $k$)
approximate probability distributions of the
cumulant estimators $\hat{\kappa}_r(k)$ for a Gaussian process with
the same power spectrum as the original test map. By comparing these
probability distributions with the cumuluant spectra of the original
image, we thus obtain an estimate of the probability that the original
image was drawn from a Gaussian ensemble. For $256 \times 256$ images 
considered here, the entire analysis requires about 20 minutes CPU time
on a Sparc Ultra workstation.

We note that, in two-dimensional applications it is usual to consider
only the domains in Fig.~\ref{domains} with $j_1, j_2 \geq 1$. The
main reason is that the domains with $j_1=0$ or $j_2=0$ contain a
mixture of coefficients of the three different two-dimensional wavelet
bases $\phi_{0,0;l_1,l_2}(x,y)$, $\zeta_{j_1,0;l_1,l_2}(x,y)$ and
$\xi_{0,j_2;l_1,l_2}(x,y)$. Domains with $j_1,j_2 \geq 1$, however,
contain only the coefficients $b_{j_1,j_2;l_1,l_2}$ of the basis
$\psi_{j_1,j_2;l_1,l_2}(x,y)$ and, as a result, statistics based on
the distribution of these wavelet coefficients are generally more
informative. Therefore, for the remainder of this paper, we will
restrict our attention to domains with $j_1, j_2 \geq 1$.

\subsection{Choosing the wavelet basis}
\label{wavechoose}

Our only remaining task is to choose the wavelet basis in which we
wish to expand our CMB map. As mentioned above, in
this paper we restrict ourselves to discrete, orthogonal,
compactly-supported wavelet bases. Nevertheless, there are still an infinite
number of such bases and so we will concentrate on only a small subset
of commonly-used wavelet bases. Our wavelet `library' consists of the
following one-dimensional wavelet bases: Haar; Daubechies 4,6,12,20;
Coiflet 2,3; Symmlet 6,8.  Several of these wavelets are plotted and
discussed by Graps (1995). We note that this library is in
no way intended to be an exhaustive set of useful wavelets, but represents
a fairly typical range of bases which could be employed. 
As mentioned above, in each case, the
corresponding two-dimensional wavelet basis is formed by taking the
tensor product of the one-dimensional basis.  

Since each wavelet basis provides a different decompostion of the
CMB test image, the resulting cumulant spectra will be different in
each case. Indeed, depending on the structure present in the test
image, some wavelet bases will be more powerful at detecting
non-Gaussianity than others. Since the analysis discussed above
requires the wavelet transform of a large number of EGR to be
calculated, it would clearly be very time-consuming to perform this
for every available wavelet basis. Therefore, for a given test image,
we wish to find a quick method by which we can determine the wavelet
basis that is most sensitive to any non-Gaussian structure present.

We begin by calculating the wavelet coefficients $b_{j_1,j_2;l_2,l_2}$
of the single test map in each of the available bases, which requires
only a couple of seconds of CPU time on a Sparc Ultra workstation.
For each basis, we then define the `normalised' wavelet coefficients 
at each value of $k$ by
\[
p_{j_1,j_2;l_2,l_2} = 
\frac{b^2_{j_1,j_2;l_2,l_2}}
{\sum_{j_1,j_2,l_1,l_2} b^2_{j_1,j_2;l_2,l_2}},
\]
where, in the denominator, the sum on the scale indices $j_1$ and $j_2$ is for all values satisfying
$2^{j_1}+2^{j_2}=k$ and the sum on position indices 
$l_1$ and $l_2$ corresponds to all wavelets coefficients lying within the
relevant scale domains. Thus, for each value of $k$,
the coefficients $p_{j_1,j_2;l_2,l_2}$ have values
between zero and unity and the sum of all the coefficients is equal to
unity. We then calculate (a multiple of) the entropy of the normalised wavelet
coefficients for each value of $k$,
\[
S = -\frac{1}{\ln N}
\sum_{j_1,j_2,l_1,l_2} p_{j_1,j_2;l_2,l_2} \ln p_{j_1,j_2;l_2,l_2}
\]
where again the sum extends over all wavelet coefficients sharing the
$k$-value under consideration and $N$ is the total number of these
coefficients. We also note
that $\lim_{p \to 0} p\ln p = 0$. The value of the sum lies
between $0$ and $-\ln N$, and 
so we have normalised $S$ so that its value always lies
between zero and unity. The entropy $S$ takes the value zero if one of the
coefficients $p_{j_1,j_2;l_2,l_2}$ 
equals unity and the rest are zero. At the other extreme,
$S$ equals unity when all the coefficients are equal.

Now, if the original CMB map is Gaussian, then, as mentioned above,
the distribution of wavelet coefficients at each $k$-value is also Gaussian.
Furthermore, for a Gaussian CMB map with structure on a wide range of
scales, we would expect that, at each scale $k$, all the wavelet basis
functions would be required, with roughly equal amplitudes, in order
to represent the image on that scale.  Hence the entropy of the
normalised wavelet coefficients should be close to its maximum value
of unity.  For a map with pronounced non-Gaussian features, however,
this is not necessarily the case. If there exists a wavelet basis that
resembles the non-Gaussian features in the map on some scale, then we
might expect the original image to be well represented by fewer basis
functions. In this case, many of the wavelet coefficients may be close
to zero, with just a few large coefficients corresponding to the basis
functions that best resemble the main features in the image. We
would therefore expect a lower value for the entropy of the normalised
wavelet coefficients. Thus, for each wavelet basis in our library, we
calculate the entropy of the wavelet coefficients of the test image at
each value of $k$. 
The optimal basis is then taken as that having the lowest entropy
for any $k$-value. For each of the test maps discussed in the
next Section, the Coiflet 2 wavelet basis was found to be optimal.
In particular, the entropy of the wavelet coefficients was 
especially low for $k=66$, which, as we shall see below, is the angular
scale on which a non-Gaussian signal is detected with the highest
significance for each test map.

\section{Application to simulated CMB maps}
\label{application}

\begin{figure}
\centerline{\epsfig{
file=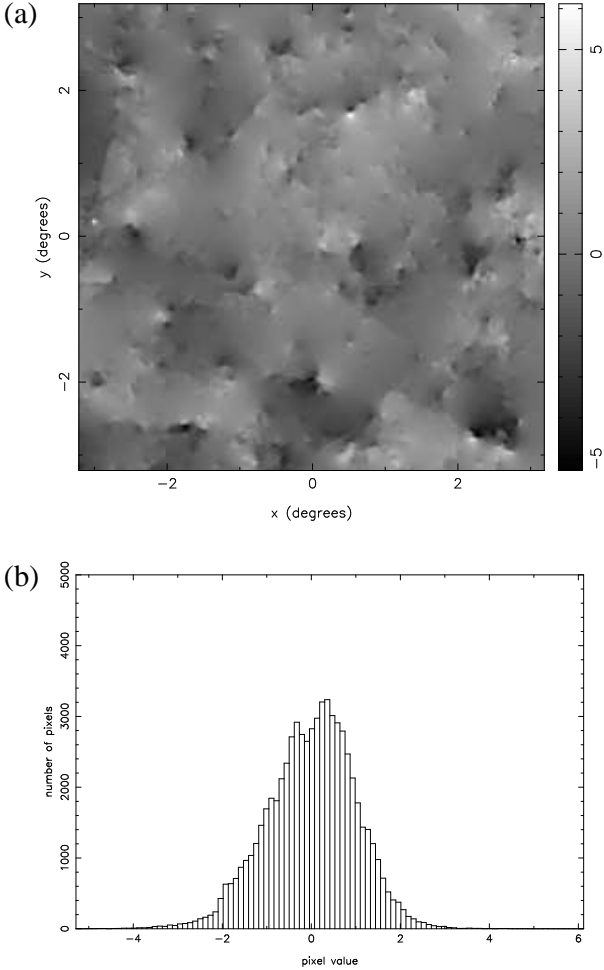,width=8cm}}
\caption{(a) A realisation of temperature fluctuations in the CMB due to
the Kaiser-Stebbins effect from cosmic strings. The map is pixelised
on a $256\times 256$ grid with a pixel size of 1.5 arcmin. 
For convenience the map
has been normalised so that its variance equals unity. (b) A
histogram of the pixel temperature distribution.}
\label{rawstrings}
\end{figure}
\begin{figure}
\centerline{\epsfig{
file=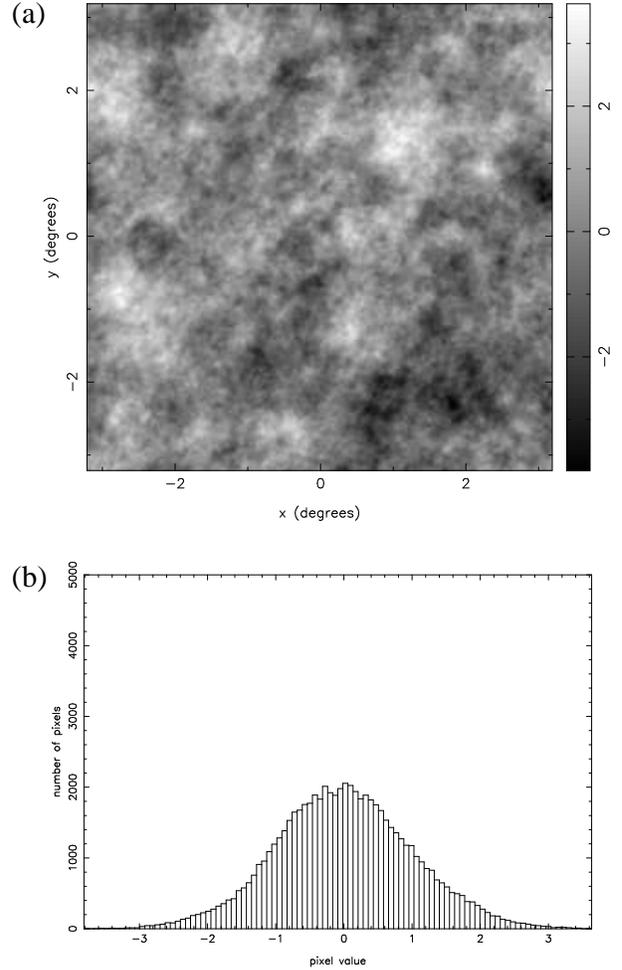,width=8cm}}
\caption{(a) A realisation of a Gaussian image with a power spectrum
identical to that of the cosmic strings map shown in
Fig.~\ref{rawstrings}(a); see
text for details. (b) A histogram of the pixel temperature distribution.}
\label{rawegi}
\end{figure}

In this Section, we apply the wavelet transform technique described
above to the detection of non-Gaussianity in simulated CMB maps.  Our
basic non-Gaussian test map is shown in Fig.~\ref{rawstrings}(a) and is a
realisation of CMB anisotropies due to the Kaiser-Stebbins
effect from cosmic strings (Bouchet, Bennett \& Stebbins 1988).
The map is pixelised on a
$256\times 256$ grid with a pixel size of 1.5 arcmin. Thus the size of
the simulated field is $6.4\times6.4$ deg$^2$. The
map has zero mean and has, for convenience, been normalised so that
its variance equals unity.

We see from Fig.~\ref{rawstrings}(a) that the strings map is extremely
non-Gaussian, possessing sharp temperature gradients and localised
hot-spots. From the scale-bar, we also note that the maximum
and minimum temperatures in the map are respectively $6.1$ and $-5.3$ times
the rms value, again indicative of a highly non-Gaussian process.
This non-Gaussianity is also apparent from the histogram of pixel
temperatures shown in Fig.~\ref{rawstrings}(b). 

In order to assess to what extent the non-Gaussian signal in the
strings map can be obscured, 
but still remain detectable, we will add to it some multiple of
the Gaussian test map (GTM) shown in Fig.~\ref{rawegi}(a). 
The GTM shown is an equivalent Gaussian realisation of the cosmic
strings map shown in Fig.~\ref{rawstrings}(a) and thus has an
identical power spectrum, but the pixel temperatures are drawn from a
Gaussian distribution. Clearly, by construction, the GTM also has
unit variance.

We could, of course, use a GTM that is not an equivalent Gaussian
realisation of the strings map. For example, we could use a
realisation of CMB fluctuations predicted from some inflationary
model. Such a map would, however, in general possess a different power
spectrum to the strings map and differences between the two could be
discerned by measuring these different power spectra. By using a GTM
that is an EGR of the strings map, we are considering the case of
optimal confusion between the non-Gaussian and Gaussian signals and
this should provide the most stringent test of the methods for detecting
non-Gaussianity. Moreover, as mentioned in the Introduction, CMB
fluctuations from realistic cosmic string cosmologies 
may well contain a Gaussian component that has 
a similar power spectrum to that of the non-Gaussian component.
(Magueijo \& Lewin 1997).

The differences between the strings map and the GTM are clear to the
eye. The GTM possesses no sharp features or extreme maxima
or minima. Furthermore, from the scale-bar, we see that
the maximum and minimum of the pixel temperature distribution lie at
3.6 and -3.8 times the rms level, as one might expect for a Gaussian
image with this number of pixels. 
The histogram of the pixel temperatures in the GTM is shown
in Fig.~\ref{rawegi}(b) and is clearly closer to a Gaussian in shape than the
corresponding histogram for the strings map shown in Fig.~\ref{rawstrings}(b).

Our test maps thus consist of the sum of the above
strings map and GTM in various proportions $(a:b)$. In order to make the
simulations more realistic, before being analysed 
the test maps are first convolved
with a 5-arcmin Gaussian beam and Gaussian pixel noise is added with
an rms level equal to one-tenth that of the convolved map. This degree
of smoothing and noise level is typical of what might be achieved
for CMB observations using the ESA Planck Surveyor satellite (Bersanelli et
al. 1996), after foreground contaminants have been removed using
the maximum-entropy separation algorithm discussed by Hobson et
al. (1998). The process of convolution smoothes out many of the
sharp features visible in the underlying cosmic strings map in
Fig.~\ref{rawstrings}(a), and the addition of pixel noise obscures
this structure still further. In each case, the resulting convolved, noisy 
test map is rescaled to have unit variance before being analysed
for any non-Gaussian signal.

As mentioned above, for a wide range of non-Gaussian-to-Gaussian
proportions $(a:b)$, the Coiflet 2 wavelet basis was found to be
optimal for detecting the non-Gaussian signal due to the strings map 
in Fig. ~\ref{rawstrings}(a). Therefore, for the remainder of this paper, all
wavelet decompositions will be in this basis. Furthermore, it was
found, for our simulations, that the fourth cumulant
spectrum, $\hat{\kappa}_4(k)$, 
was the most sensitive to the presence of any non-Gaussian signal,
and so we will plot only this spectrum for each test map. 

\subsection{The Gaussian test map}
\label{puregmap}

\begin{figure}
\centerline{\epsfig{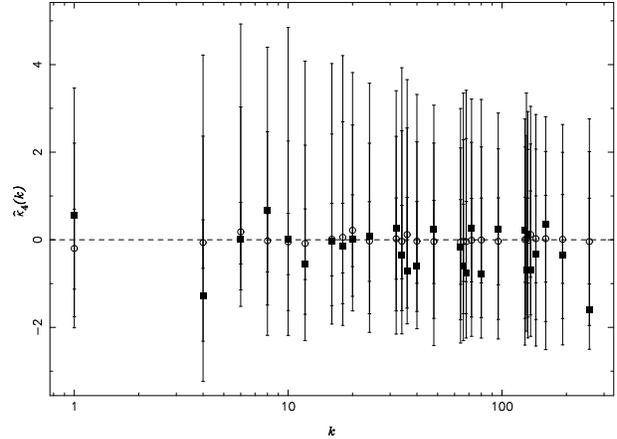}}
\caption{The $\hat{\kappa}_4(k)$ spectrum (solid squares) of the
wavelet coefficients of the test map consisting of the sum of the
cosmic strings map shown in Fig.~\ref{rawstrings}(a) and the
Gaussian test map shown in Fig.~\ref{rawegi}(a) in the
proportions $(0:1)$, after convolution with a 5-arcmin Gaussian beam
and the addition of Gaussian pixel noise with rms equal to one-tenth
that of the convolved map. The open circles show the mean
$\hat{\kappa}_4(k)$ spectrum obtained from 5000 equivalent Gaussian
realisations of the test
map (see text for details) and the error bars show the 68, 95 and 99
per cent limits of the resulting $\hat{\kappa}_4$ distribution at each
value of $k$. For presentational purposes, the spectrum been
normalised so that, at each value of $k$, the variance of the
$\hat{\kappa}_4$ distribution obtained from the 5000 EGR is equal to
unity. The point at $k=1$ is {\em not} calculated from the wavelet
coefficients, but instead shows the values of $\hat{\kappa}_4$
obtained directly from the pixel temperature distributions in the test
map and the 5000 EGR.}
\label{kspec1}
\end{figure}

We begin by assuming the CMB fluctuations to consist only of the
Gaussian map shown in Fig.~\ref{rawegi}(a).  Thus,
formally, our initial test map consists of the sum of the cosmic
strings map and the GTM in the proportions $(0:1)$ . This is then
convolved with a 5-arcmin beam and pixel noise is added as discussed
above.  Applying the wavelet non-Gaussianity test to this image
provides a useful check that a detection of non-Gaussianity is
not obtained for a purely Gaussian signal.  The resulting
$\hat{\kappa}_4(k)$ cumulant spectrum of the wavelet coefficients 
of the test map is plotted as
the solid squares in Fig.~\ref{kspec1}.  As mentioned in Section
\ref{wtstats}, the same non-Gaussianity test is then also applied to
5000 equivalent Gaussian realisations (EGR) of the test map.
In Fig.~\ref{kspec1}, the mean
value of $\hat{\kappa}_4(k)$ obtained from the 5000 EGR is shown by
the open circles, which all lie close to zero, as expected for a
Gaussian process.  At each value of $k$, the error bars show the 68,
95 and 99 per cent limits of the $\hat{\kappa}_4$ distribution
obtained from the 5000 EGR. For presentational purposes, the cumulant
spectrum has in fact been normalised so that, at each value of $k$,
the variance of the $\hat{\kappa}_4$ distribution obtained from the
5000 EGR is equal to unity. Thus the significance of any detection of
non-Gaussianity can be read off directly from the scale on the
vertical axis of the plot. As might be expected in this case, however, the
$\hat{\kappa}_4(k)$ spectrum of the test map lies well within the
2-sigma limits of the Gaussian ensemble at all values of $k$.

As discussed in Section \ref{wtstats}, since we ignore domains of the
wavelet transform with $j_1$ or $j_2$ equal to zero, we see from
Fig.~\ref{domains} that the $\hat{\kappa}_4(k)$ spectrum of the
wavelet coefficients should start at $k=4$.  The point plotted at
$k=1$ in Fig.~\ref{kspec1} is {\em not} in fact calculated from the
wavelet coefficients, but instead shows the values of $\hat{\kappa}_4$
obtained directly from the pixel temperature distributions in the test
map and the 5000 EGR. This point therefore represents a
non-Gaussianity test similar to that proposed by Scaramella \&
Vittorio (1991), in which the skewness and kurtosis of the pixel
temperature distributions were calculated. The point is included to
provide an immediate comparison between methods based on the
statistics of the wavelet coefficients and those based directly on the
statistics of the pixel temperature distribution.

\subsection{The cosmic strings map}
\label{stringmap}

\begin{figure}
\centerline{\epsfig{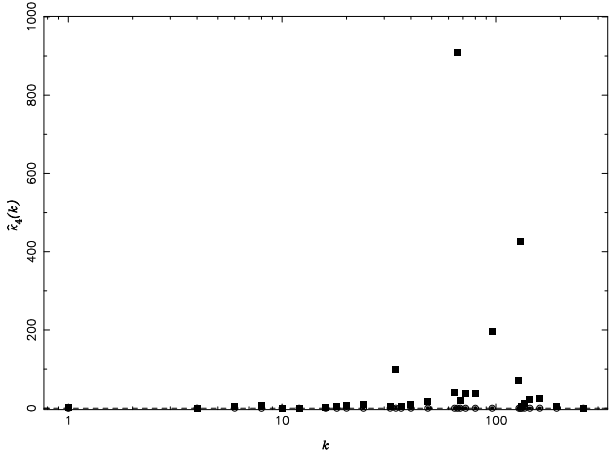}}
\caption{As in Fig.~\ref{kspec1}, but for the test map 
with non-Gaussian: Gaussian proportions $(1:0)$.}
\label{kspec2}
\end{figure} 

We may repeat the above analysis for the opposite extreme in which the
CMB fluctuations are assumed to consist only of the cosmic strings
contribution (so that the non-Gaussian:Gaussian proportions are
formally $(1:0)$). As before, the map is then
convolved with a 5-arcmin beam and pixel noise is added.
The resulting $\hat{\kappa}_4(k)$ spectrum is shown in Fig.~\ref{kspec2}.
We see from the figure that unambiguous detections of
non-Gaussianity are obtained at numerous values of $k$, the most
spectacular being a 910-sigma detection at $k=66$. As discussed in
Section \ref{wavechoose}, the entropy of the wavelet coefficients 
for the Coiflet 2 basis was lowest 
in the domains with $k=66$, and so we might have expected these
domains to provide the most significant detection of non-Gaussianity.

Indeed, since the detection of non-Gaussianity is so large, it is not
possible to discern the error bars denoting the confidence limits of
the $\hat{\kappa}_4$ distributions for the 5000 EGR (although they are
plotted). We merely note that for the $k=1$ point (calculated directly
from the pixel temperatures), the significance of the non-Gaussian
detection is only 1.1-sigma.  Thus, in this case we find that, by
analysing the wavelet coefficients of the image in separate
$k$-domains, we obtain a vastly improved detection of non-Gaussianity
as compared with analysing the pixel temperature distribution
directly.

\subsection{Mixed cosmic strings/Gaussian maps}
\label{mixedmaps}

\begin{figure*}
\centerline{\epsfig{
file=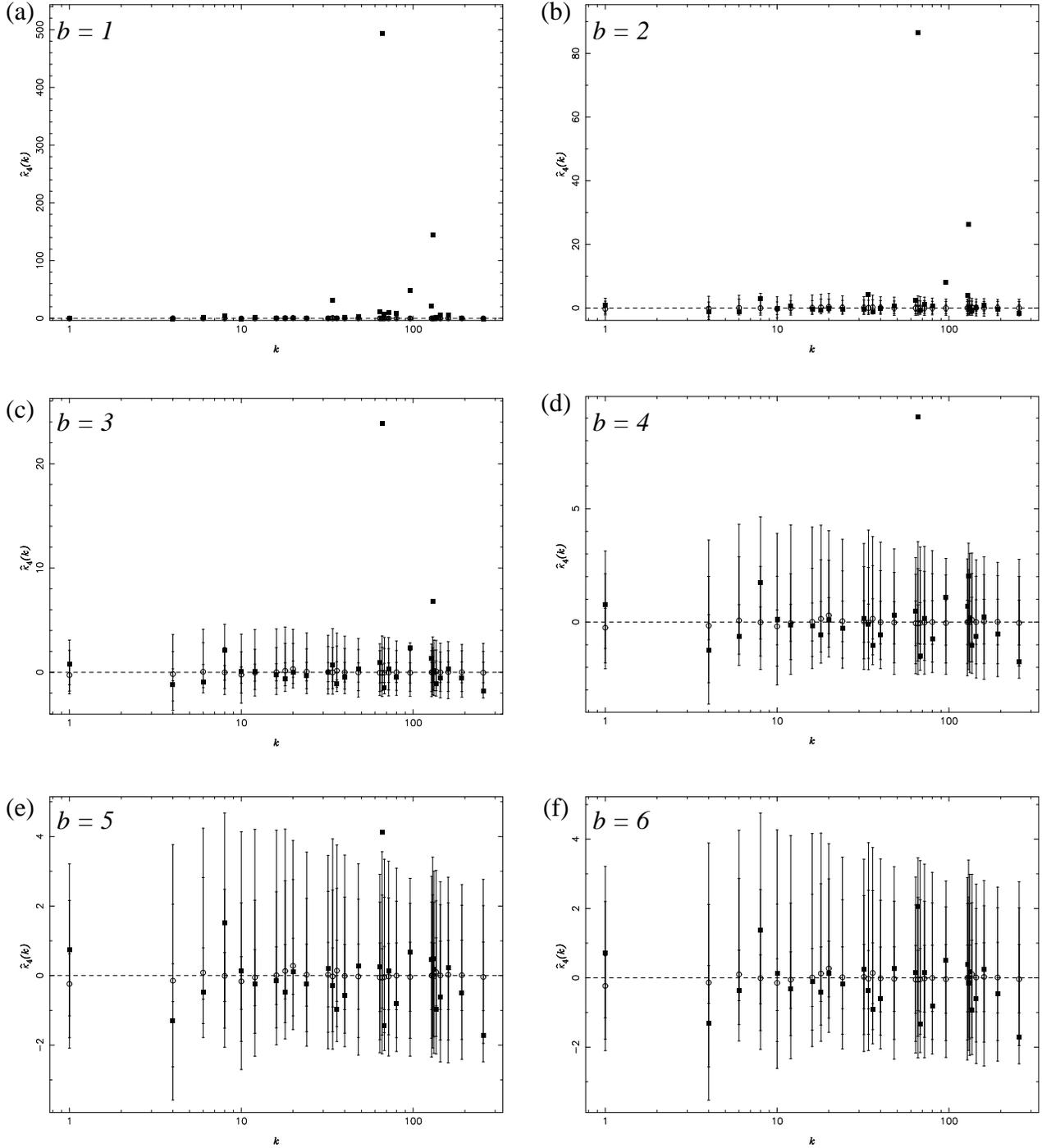,width=16.8cm}}
\caption{As in Fig.~\ref{kspec1}, but for test maps 
with non-Gaussian: Gaussian proportions $(1:b)$.}
\label{kspec3}
\end{figure*}

Now we have analysed separately the cosmic strings map in
Fig.~\ref{rawstrings}(a) and the GTM in Fig.~\ref{rawegi}(a) (after
convolution and the addition of pixel noise), we may repeat the
analysis for test maps consisting of the sum
of the cosmic strings map and GTM in different proportions $(1:b)$.
It is clear that when $b \gg 1$ 
the resulting $\hat{\kappa}_4(k)$ spectrum of the wavelet coefficients
will tend to that shown in Fig.~\ref{kspec1}. Similarly, if $b \ll 1$
the $\hat{\kappa}_4(k)$ spectrum will lie close to that in
Fig.~\ref{kspec2}. In this Section, our aim is to find the value of
$b$ at which the detection of non-Gaussianity becomes marginal.
We reiterate that, in the following analysis, each test map is
convolved and has noise added (as discussed above) before applying
the non-Gaussianity test.

In Fig.~\ref{kspec3} we plot the $\hat{\kappa}_4(k)$ spectra for test
maps with $b=1,2,3,4,5,6$.  We see from the figure that significant
non-Gaussian detections are obtained for $b \leq 4$. We note that, in
each case, the largest detection of non-Gaussianity occurs at $k=66$,
as expected from the entropy analysis discussed in Section
\ref{wavechoose}.  The significance of the largest detections for
$b=1,2,3,4$ is 495-, 88-, 24- and 9-sigma respectively.  For $b=5$,
the detection of non-Gaussianity becomes marginal. The largest signal
again occurs at $k=66$, but the significance is only 4-sigma. For
$b=6$, no detection is possible and, as expected, the
$\hat{\kappa}_4(k)$ spectrum closely resembles that shown in
Fig.~\ref{kspec2} for the case when no non-Gaussian signal is
present.

For all values of $b$, however, we note that for the $k=1$ point,
which is calculated directly from the pixel temperatures, the value of
$\hat{\kappa}_4$ for the test map lies within the 95 percent
confidence limits of the of $\hat{\kappa}_4$ distribution obtained
from the 5000 EGR. Thus, we see the statistic based on the pixel
temperature distribution is far less sensitive to the presence of a
non-Gaussian signal than those based on the wavelet coefficients of
the image.

The performance of the wavelet non-Gaussianity test is encouraging. In
particular, we must remember that the parameter $b$ represents the
ratio of the rms values of the GTM and cosmic strings maps that
constitute the test image. The ratio of the power in the Gaussian and
non-Gaussian contributions is given by $b^2$ and so a marginal
detection is still possible when the power due to the GTM is 25 times
that due to the cosmic strings map. We note that, since the GTM possess
an identical power spectrum to the cosmic strings map, the ratio
of the power in the Gaussian and non-Gaussian contributions is equal
to $b^2$ on all angular scales. Thus, the detection of
non-Gaussianity is not the result of the presence of excess power in
the cosmic strings map over some range of scales.

\begin{figure}
\centerline{\epsfig{
file=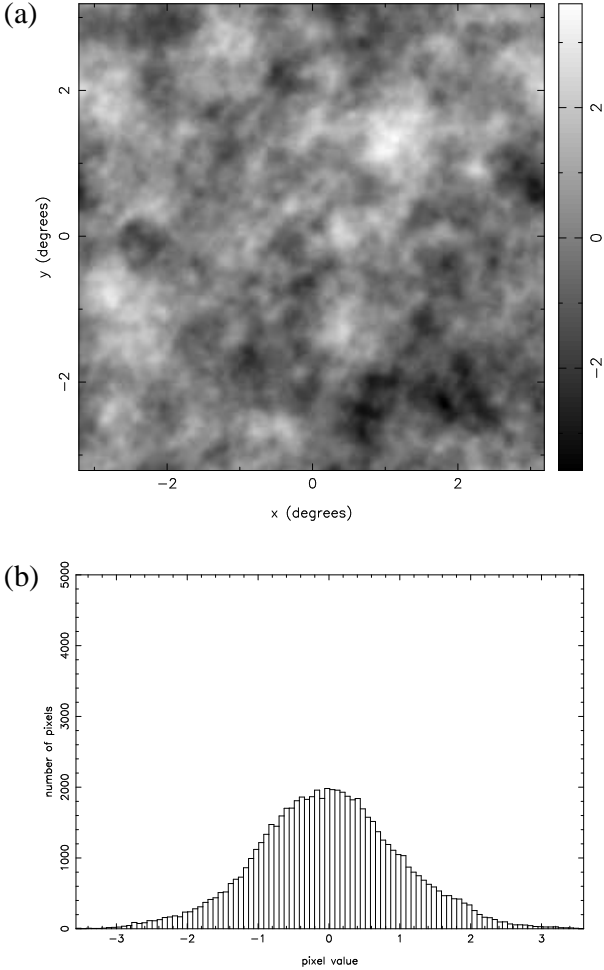,width=8cm}}
\caption{(a) The test map containing the cosmic strings map
shown in Fig.~\ref{rawstrings}(a) and the EGR shown on
Fig.~\ref{rawegi}(a) in the proportions $(1:5)$. The map has been convolved
with a 5-arcmin Gaussian beam and Gaussian pixel noise has been added
with an rms equal to one-tenth that of the convolved CMB map.
For convenience the map has been normalised to that its variance 
equals unity. (b) A histogram of the pixel temperature distribution.}
\label{prop15}
\end{figure}

For illustration purposes, in Fig.~\ref{prop15}(a) we plot the $b=5$
test map, after convolution and the addition of pixel noise. Although
the wavelet non-Gaussianity test provided a marginal detection, as
shown in Fig.~\ref{kspec3}(e), we see that, at least by eye, the
non-Gaussian signal from the cosmic strings map in
Fig.~\ref{rawstrings}(a) is completely obscured by the signal due to
the GTM in Fig.~\ref{rawegi}(a). Moreover, convolution with a 5-arcmin
Gaussian beam and the addition of pixel noise have further diluted the
non-Gaussian signal. The histogram is pixel temperatures is shown in
Fig.~\ref{prop15}(b) and is clearly close to Gaussian.

\subsection{The cosmic strings map with enhanced pixel noise }
\label{hinoise}

As a final example, we consider again the case in which the CMB signal
consists only of the cosmic strings map, convolved to 5-arcmin
resolution (as in Section \ref{stringmap}), but in which the level of
instrumental pixel noise is increased such that its rms value is equal
to that of the convolved CMB map. This example is included for two
reasons. Firstly, it has traditionally been pixel noise that presented
the greatest obstacle to the detection of non-Gaussianity in the CMB,
and this is still the case for the 4-year COBE data. Although the
observational strategy for our simulated data is very different from
that used for the COBE observations, the information content of the two
data-sets is comparable.  The $6.4\times 6.4$ deg$^2$ map used here
contains approximately 6000 5-arcmin beams, which is similar to the
number of pixels in standard resolution COBE observations, although
this somewhat larger than the number of independent COBE beams.
The second reason for
including this test image is to investigate 
the abilities of the wavelet technique when
the main contaminant does not possess a power spectrum that is
identical in shape to that of the non-Gaussian signal, but instead has
a flat white-noise power spectrum.

\begin{figure}
\centerline{\epsfig{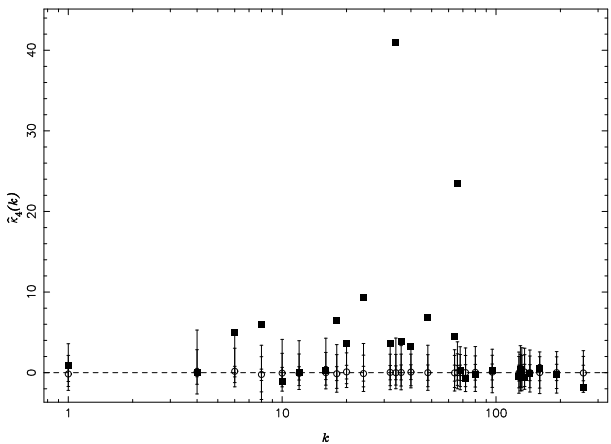}}
\caption{As in Fig.~\ref{kspec1}, but for the test map 
with non-Gaussian: Gaussian proportions $(1:0)$ to which Gaussian pixel noise
has been added with an rms value equal to that of the convolved CMB map.}
\label{kspecnoise}
\end{figure}

The resulting $\hat{\kappa}_4(k)$ spectrum for this test map is shown
in Fig.~\ref{kspecnoise}. We see from the figure that the most
significant detection of non-Gaussianity is 42-sigma at $k=34$,
accompanied by a 24-sigma detection at $k=66$ and several 5--10-sigma
detections at other $k$-values.  It is interesting to note that the
most significant detection is much smaller than the 495-sigma
detection obtained in Fig.~\ref{kspec3}(a) and occurs at a lower value
of $k$. For Fig.~\ref{kspec3}(a), the main contaminant also had an rms equal
to that of the cosmic strings map, but possessed an identical power
spectrum. It is clear therefore that the wavelet technique is less
effective at detecting the non-Gaussian signal due to the cosmic
strings map when the main contaminant has a flat white-noise power
spectrum. This is in fact to be expected, since the non-Gaussian
signal in the strings map comes predominantly from sharp temperature
gradients, which occur on small scales. As compared to the cosmic
strings map, a pixel noise contaminant with the same rms value has
relatively more power on small scales, and is therefore more effective
at obscuring the non-Gaussian signal. By comparing
Figs~\ref{kspec3}(a) and \ref{kspecnoise} more closely, this behaviour
becomes more apparent. We see that in both cases a $\sim 40$-sigma
detection is obtained at $k=34$, but the more significant detections
in Fig.~\ref{kspec3}(a), which occur at higher $k$-values, have been
much reduced in Fig.~\ref{kspecnoise}.


\section{Minkowski functionals analysis}
\label{minkfunc}

In the previous Section, we found that the wavelet non-Gaussianity
test was far more sensitive to the presence of a non-Gaussian signal
than statistics based simply on the pixel temperature distribution of
the CMB map. In order to compare the wavelet approach with
other methods for detecting non-Gaussianity, in this
Section we reanalyse the above test maps
using an approach based on the topology of the CMB fluctuations.

The approach we adopt here requires the calculation of three 
{\em Minkowski functionals} for each test map.
Minkowski functionals have been applied to CMB data
in several forms previously. Schmalzing \& Gorski (1998) derive the
Minkowski functionals for a spherical sky and apply their results to
the COBE four year data. Winitzki \& Kosowsky (1998) give a detailed
derivation of the expected form of the Minkowski functionals for a
flat Euclidean space, taking into account boundary conditions, pixel
noise and pixel size and shape. The reader is refered there for a more
theoretical overview.

We opt here for an approach that is as simple as possible and use the three
two-dimensional Minkowski functionals on square pixels in flat
space. This approach is sufficient for the comparison desired in this
paper. The three Minkowski functionals are the surface area, perimeter
and Euler characteristic of an excursion region. The
excursion region $R(T)$ is taken as the region of the map above a certain
threshold temperature $T$, which is varied between the minimum and
maximum values in the test map. The Minkowski functionals are
therefore functions of the threshold temperature $T$.
In two dimensions, the three functionals 
completely describe the statistical properties of a
distribution. In terms of integral geometry they are given by 
\begin{eqnarray*}
M_0(T) & = & \int_{R(T)} {\rm d}S,\\
M_1(T) & = & \int_{\partial R(T)} {\rm d}l,\\
M_2(T) & = & \int_{\partial R(T)} {{\rm d}l\over{r}} .
\end{eqnarray*}
where $\partial R(T)$ is the boundary of the excursion region $R(T)$
at the threshold temperature $T$. The differentials
${\rm d}S$ and ${\rm d}l$ denote respectively the elements of area and 
of length along the boundary, and $r$ is the radius of curvature of the 
boundary. As an example, a circular disc of radius $a$ would
have $M_0=\pi a^2$, $M_1=2\pi a$ and $M_2=2\pi$.

The surface area and perimeter of the excursion region in a pixelised
map are trivial to calculate but the Euler characteristic requires a
little explanation. It should be noted that, as defined here, the
third Minkowski functional is equal to $2\pi G$ where $G$ is the genus
of the excursion region.  The genus is defined as the number of
isolated holes (regions below the threshold) minus the number of
islands (regions above the threshold) within the map.  Using the
approach advocated by Melott et al. (1989), based upon the three
dimensional genus calculation of Hamilton, Gott \& Weinberg (1986), we
calculate the genus of the two-dimensional map by looking at the angle
deficits of the vertices. Using this method it is also straightfoward
to calculate the three Minkowski functionals simultaneously.

This approach to genus has been implemented previously and applied to
the COBE data by Smoot et al. (1994), Kogut et al. (1996)
and Colley, Gott \& Park (1996). Gott et al. (1990) also looked
at the use of the perimeter of the excursion regions but did not apply
this analysis to any data. Theoretical predictions from defect
theories have also been tested using the genus algorithm (see e.g.
Avelino et al. (1998)) but the full Minkowski functional set is
yet to be applied to a full CMB data set to constrain any cosmological
parameters or defect theories. This has been carried out for various Large
Scale Structure data sets (see e.g. Kerscher et al. 1997) with
promising results.

\subsection{Application to simulated CMB maps}

\begin{figure*}
\centerline{\epsfig{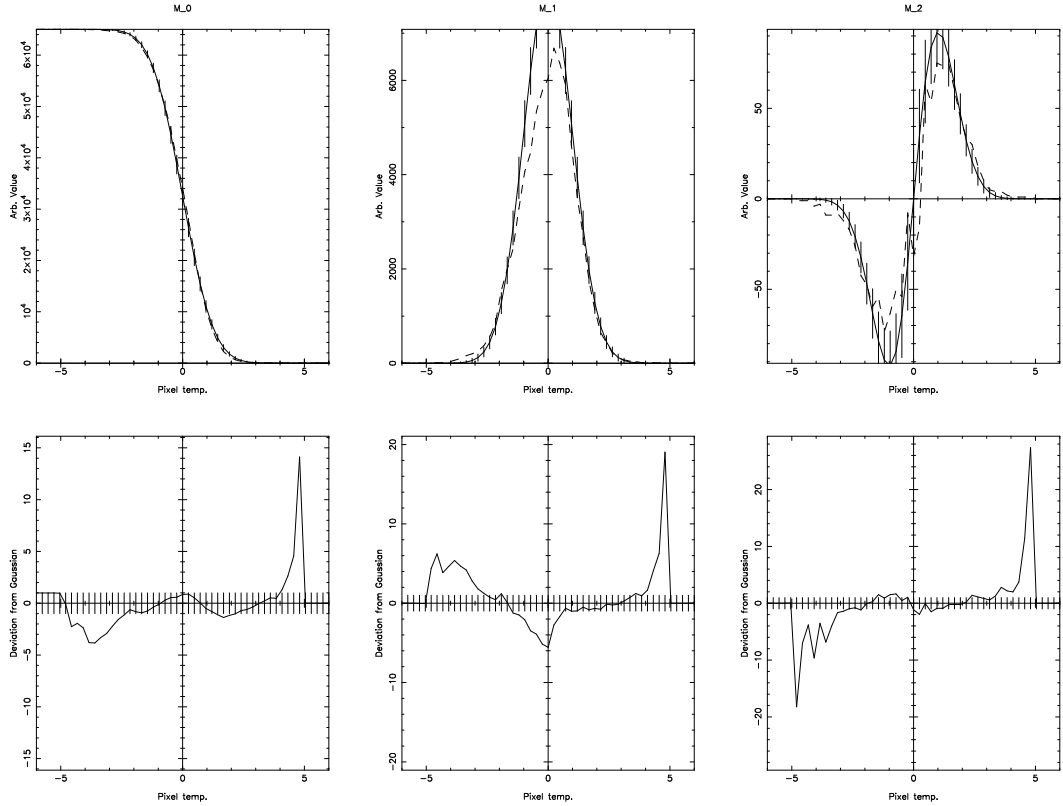}}
\caption{Minkowski functionals analysis of the test map with
non-Gaussian: Gaussian proportions $(1,0)$.
{\em Top panels}: the 
three Minkowski functionals of the test map (dashed lines) calculated
at 50 separate threshold temperatures between the minimum and maximum
of the test map pixel temperature distribution. The solid lines show
the mean Minkowski functionals calculated from 5000 EGR and the error
bars denote the
one-sigma limits of the probability distributions of the
functionals at each temperature step.
{\em Bottom panels}: 
the difference between the Minkowski functionals of the test map and
mean Minkowski functionals of the 5000 EGR. At each temperature step,
the plots have been normalised so that the one-sigma points of the 
probability distributions obtained from the 5000 EGR are equal to unity.}
\label{minkfig1}
\end{figure*}

The three Minkowski functionals were calculated for the simulated CMB
maps analysed in the previous Section.  As discussed above, these test
maps consist of the strings maps in Fig.~\ref{rawstrings}(a) and the
EGR in Fig~\ref{rawegi}(a) in proportions $(1:b)$, where $b$ can vary.  Before
being analysed, the test maps are first convolved with a 5-arcmin
Gaussian beam and Gaussian pixel noise is added with an rms equal to
one-tenth the rms of the convolved map.
In each case the temperature range of the test map is divided into 50
steps and, at each temperature step, the three Minkowski functionals of the
corresponding excursion region $R(T)$ are calculated.
Following the procedure adopted in the wavelets non-Gaussianity test,
in each case the three Minkowski functional are also calculated 
in an analogous manner for
5000 EGR so that approximate probability distributions are obtained
for the values of the functionals at each temperature step.

\begin{figure*}
\centerline{\epsfig{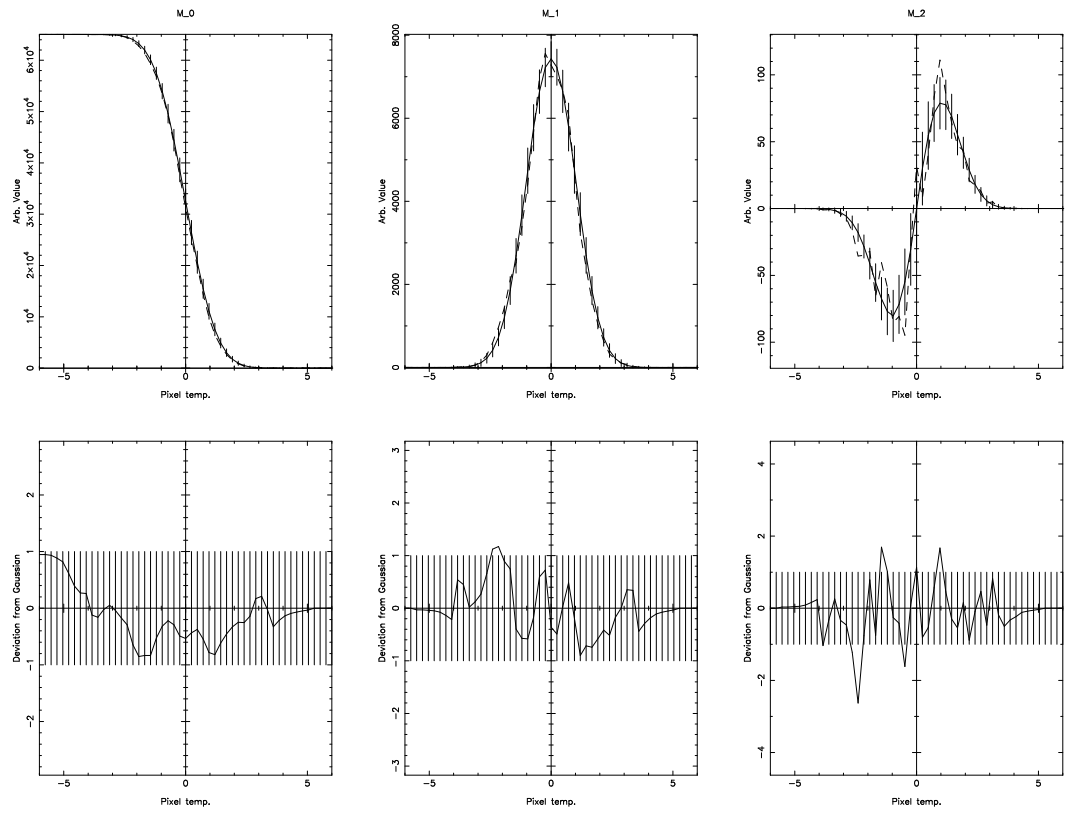}}
\caption{As in Fig.~\ref{minkfig1}, but for the test map with
non-Gaussian: Gaussian proportions $(1:1)$.}
\label{minkfig2}
\end{figure*}
\begin{figure*}
\centerline{\epsfig{
file=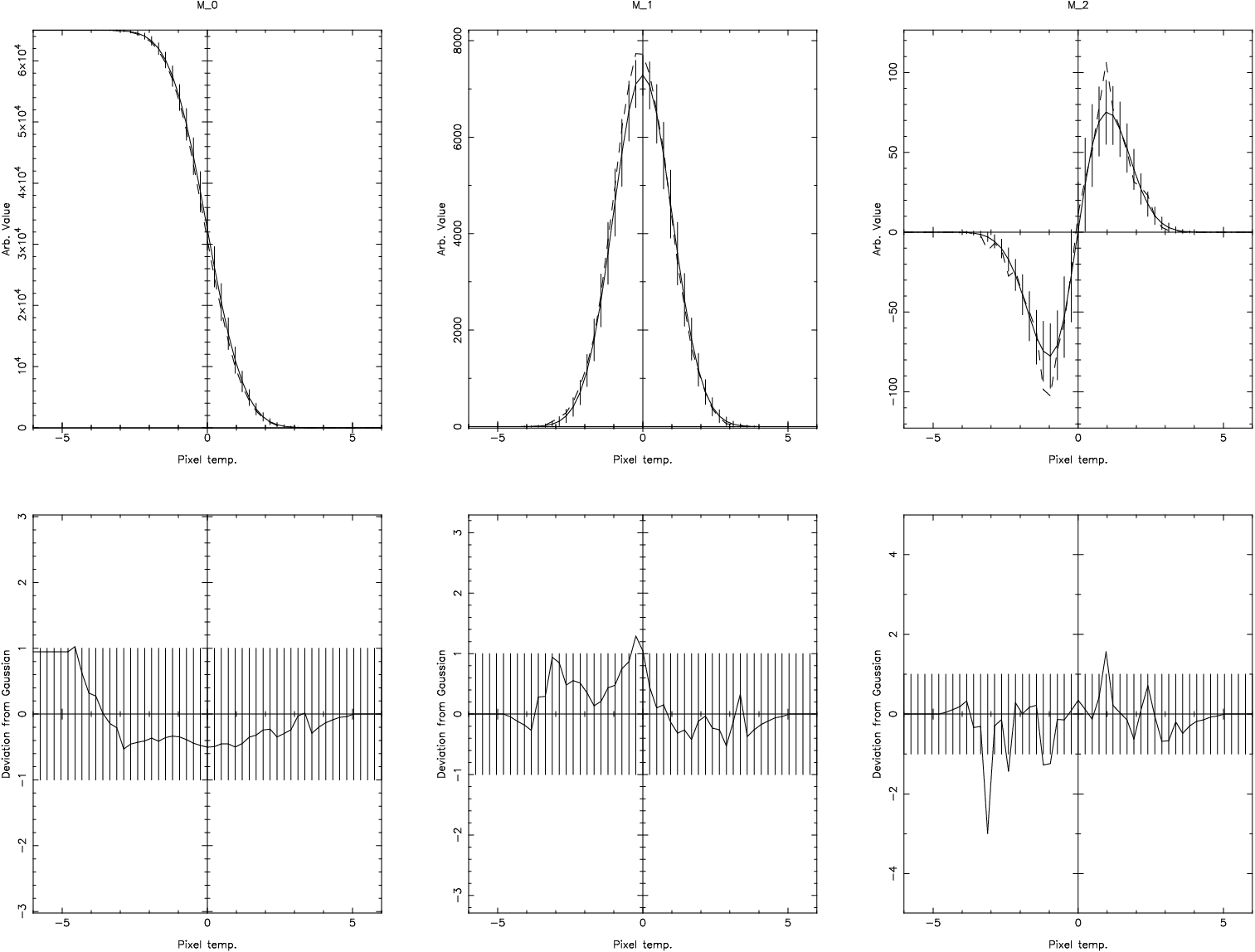,angle=270,width=14cm}}
\caption{As in Fig.~\ref{minkfig1}, but for the test map with
non-Gaussian: Gaussian proportions $(1:2)$.}
\label{minkfig3}
\end{figure*}
\begin{figure*}
\centerline{\epsfig{
file=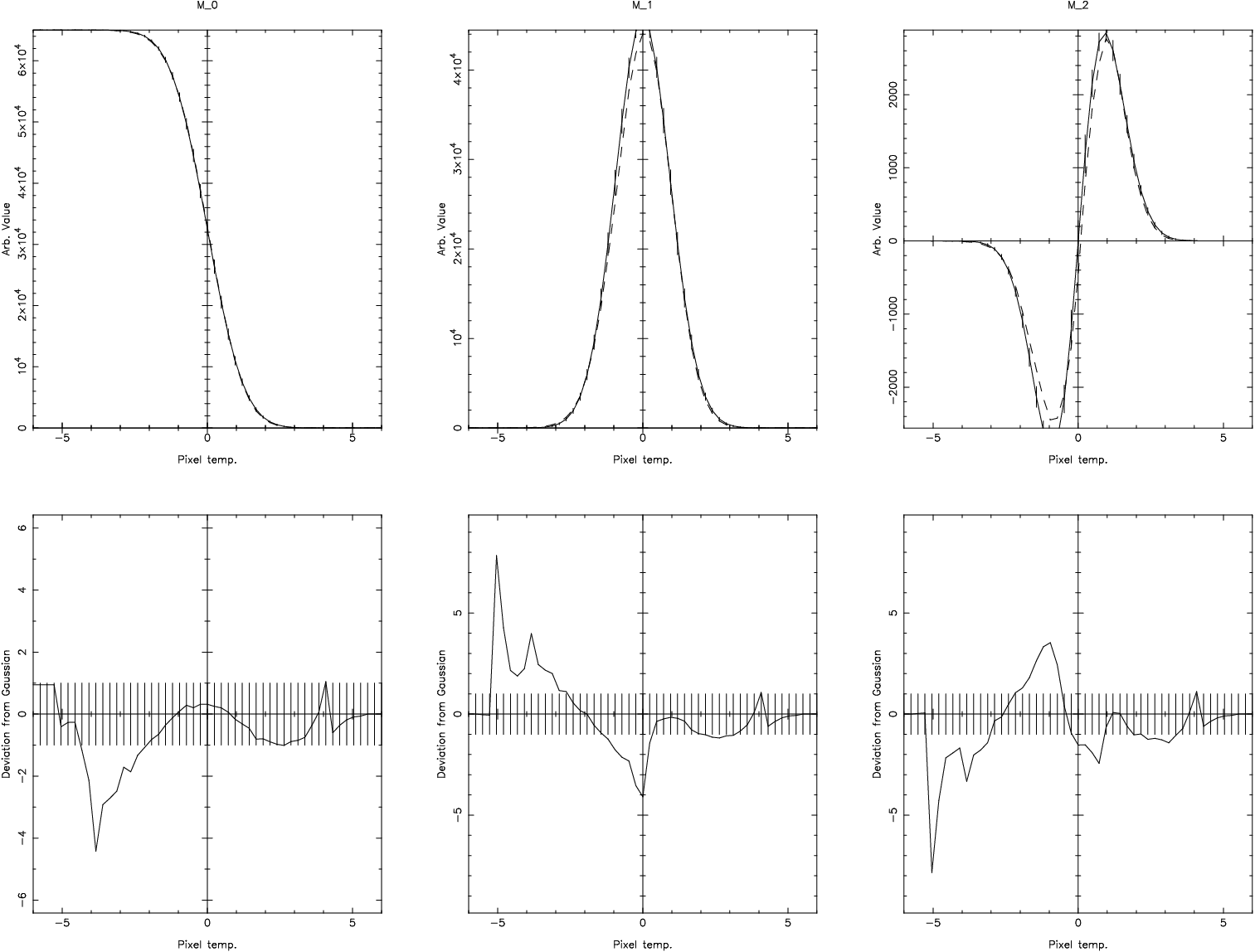,angle=270,width=14cm}}
\caption{As in Fig.~\ref{minkfig1}, but for the test map with
non-Gaussian: Gaussian proportions $(1:0)$ to which Gaussian pixel noise
has been added with an rms value equal to that of the convolved CMB map.}
\label{minkfig4}
\end{figure*}

The results for the cases $b=0$, 1 and 2 are
shown in Figures~\ref{minkfig1}--\ref{minkfig3}. 
In the top three panels of each figure, we plot the
Minkowski functionals of the test map (dashed lines)
and the mean Minkowski functional for the 5000 EGR (solid lines).
The error bars in each panel correspond to the
one-sigma limits of the probability distributions of the values of the
functionals at each temperature step, as obtained from the 5000 EGR.
In order to estimate the significance of any non-Gaussian detection,
in the bottom three panels of each figure we plot 
the difference between the Minkowski functionals of the test map and
mean Minkowski functionals of the 5000 EGR. At each temperature step,
the plots have been normalised so that the one-sigma points of the 
probability distributions obtained from the 5000 EGR are equal to unity.
Thus, the significance of any non-Gaussian detection may be read-off
directly from the vertical axis of each plot.

In the $b=0$ case (for which the test image is simply the convolved strings
map with noise added), we see from the bottom three panels in
Fig.~\ref{minkfig1} that
significant detections of non-Gaussianity are obtained for all three
Minkowski functionals. For low and high values of $T$, the
corresponding excursions region of the test map are significantly different
from those in a typical EGR and all three functionals detect some
non-Gaussianity. Indeed the $M_0$, $M_1$ and $M_2$ 
functionals have detections of 14, 20 and 24-sigma respectively
at high positive threshold temperatures.
For the $M_1$ functional, we also see that, when the
threshold temperature $T$ is close to zero (the mean of the maps), the
excursion region $R(T)$ of the test map has a significantly smaller
perimeter than for a typical EGR and a 6-sigma detection of
non-Gaussianity is obtained. 

This behaviour is easily understood by examining the structure present
in the cosmic strings map shown in Fig.~\ref{rawstrings}. 
The strings map possesses a
great deal of structure for threshold temperatures near the minimum
and maximum of the pixel temperature distribution.  Even after
convolution with the 5-arcmin beam, there still exist several sharply
defined hot-spots and cold-spots that lie outside the typical 3-sigma
range of the corresponding EGRs, and hence lead to large detections of
non-Gaussianity.  As noted earlier, the range of values in the
(unconvolved) strings map is 6.1 to -5.3, and so at its extremes the
temperature distribution is slightly skewed to positive values, which
leads to more pronounced detections at high temperatures. Thus, as
might be expected, the Minkowski functional analysis is powerful at
detecting non-Gaussianity due to points in the strings map temperature
distribution that lie outside the corresponding distribution of a
typical EGR.
For threshold temperatures near zero, however, 
the strings map is quite structureless,
possessing large plane areas at constant temperature. Thus, the
perimeter of a correpsonding excursion region will be quite small in
comparison to an EGR and leads to the 6-sigma detection in the
$M_1$ functional.

From Figs~\ref{minkfig2} \& \ref{minkfig3}, however, we see that the results
for $b=1$ and $b=2$ are somewhat disappointing. None of the Minkowski
functionals have produced a significant detection of
non-Gaussianity. Very few points lie outside the one-sigma limits
obtained from the 5000 EGR and, even for $b=1$, the most significant
deviation is only 2.5-sigma and must certainly not be considered as a
robust detection of non-Gaussianity.  Thus, we find that, in the
presence of superposed Gaussian signals, the Minkowski functionals are
far less sensitive than the wavelet technique to the presence of the
non-Gaussian signal due to the cosmic strings map.

Finally, we perform the Minkowski functionals analysis
on the enhanced noise map discussed in Section \ref{hinoise}.
The results are shown in Fig.~\ref{minkfig4}.
In contrast to the wavelets technique, we find that the Minkowski
functionals are more sensitive to non-Gaussianity in the strings map
when the main contaminant is pixel noise as opposed to an EGR. We see
that the $M_2$ and $M_3$ functionals provide $\sim$8-sigma detections
of non-Gaussianity at lower limit of the pixel temperature distribution.

\section{Discussion and conclusions}
\label{conc}

We have investigated the use of wavelet analysis in detecting
non-Gaussianity in the CMB. Our test images consist of non-Gaussian
CMB fluctuations from the Kaiser-Stebbins effect due to cosmic strings
plus some multiple of a Gaussian map with a power spectrum identical
to the cosmic strings map.  Before being analysed, the test images are
convolved with a 5-arcmin Gaussian beam and Gaussian pixel noise is
added with an rms equal to one-tenth that of the convolved test
map. Using the wavelet technique, we find that the non-Gaussian signal
can still be detected even when the superposed Gaussian map has an rms
equal to 5 times that of the underlying cosmic strings map.  However,
statistics based directly on the pixel temperature distribution were
unable to detect the non-Gaussian signal even in the absence of a
superposed Gaussian map.  The wavelet technique also produced clear
detections of non-Gaussianity in the case where the cosmic strings map
was contaminated by Gaussian pixel noise with the same rms value; in
this regime standard methods were once again unsuccessful.

We also find that the wavelet technique
outperforms methods based on the calculation of Minkowski functionals.
Although the Minkowski functional approach yields a significant
detection of non-Gaussianity for the case in which no Gaussian map is
superposed, we find that, once a Gaussian signal with a equal rms
value is added, no significant detection is obtained. However, the Minkowski
functional technique did yield reasonable detections of
non-Gaussianity in the high pixel noise regime.
One possible way to improve the results from the Minkowski functional
analysis may be to introduce into the method some notion of identifying 
localised structure, which is inherent in the wavelet technique.
By looking at the distribution of Minkowski functionals as a function of
position across the map it may be possible to make more significant
detections of non-Gaussianity. Such a procedure has been investigated
by Novikov, Feldman and Shandarin (1998), who use `partial' Minkowski
functionals to analyse the COBE 4-year data with promising results. 

For both the wavelet and Minkowski functional analyses, we have quoted
the significance of any detection of non-Gaussianity simply as the
largest single deviation observed. This is, of course, the most
conservative assumption that we can make, and corresponds to the
situation in which a high degree of correlation exists bewteen the
points in the $\hat{\kappa}_4(k)$ spectrum or between the values of
the Minkowski functionals at different threshold temperatures.  Thus,
the significances quoted in this paper should properly be considered
as lower limits. Clearly, if the points were independent, then we may
simply multiply together the individual significances to obtain a
greatly enhanced detection of non-Gaussianity. In order to obtain a
value for the true level of significance for the wavelet or Minkowski
functional approach, it is necessary to derive expressions for the
correlations that exist between the individual points in each case. In
particular, one would like to define some form of likelihood function
that includes these correlations. We have not carried out such a
procedure here, but will investigate these matters more fully in a
forthcoming paper.

Finally, we mention how the current wavelet non-Gaussianity test might
be further improved.  The wavelet transform applied here decomposes
the two-dimensional CMB temperature fluctuations into a
two-dimensional wavelet basis. The bases used in this paper were
constructed from the direct (tensor) product of existing commonly-used
one-dimensional wavelet bases (see Section \ref{wave2d}).  The main
disadvantage of this approach is the mixing of structure with
different scales $j_1$ and $j_2$ in the horizontal and vertical
directions. As pointed out by Mallat (1989), however, it is in fact
possible to use one-dimensional wavelet bases to construct
two-dimensional bases that do not mix scales and can be described in
terms of a single scale index $j$.  At each scale level $j$, these
bases are given by
\begin{eqnarray*}
\phi_{j;l_1,l_2}(x,y) & = & \phi_{j,l_1}(x)\phi_{j,l_2}(y) \\
\psi^{\rm H}_{j;l_1,l_2}(x,y) & = & \psi_{j,l_1}(x)\phi_{j,l_2}(y) \\
\psi^{\rm V}_{j;l_1,l_2}(x,y) & = & \phi_{j,l_1}(x)\psi_{j,l_2}(y) \\
\psi^{\rm D}_{j;l_1,l_2}(x,y) & = & \psi_{j,l_1}(x)\psi_{j,l_2}(y).
\end{eqnarray*}
The $\phi_{j;l_1,l_2}$ wavelet is simply an averaging function at the
$j$th level, while the other three wavelets correspond to structure at
the $j$ scale level in the horizontal, vertical and diagonal
directions in the image. Since such a basis does not mix structure on
different scales, we might expect that it would provide wavelet
decomposition of the CMB map that is more sensitive to the presence of
any non-Gaussian signal. The application of this form of
two-dimensional wavelet basis to the detection of non-Gaussianity in
the CMB will be discussed in a forthcoming paper. 

Even with the above modification, however, the wavelet non-Gaussianity
test presented here is still restricted to flat two-dimensional
images. Thus, in the context of the CMB, it can only be applied
individually to small patches of sky for which the curvature of the
sky is not significant. In order to apply the method to all-sky CMB
maps, such as the COBE 4-year map or forthcoming maps from the MAP and
Planck Surveyor missions, it is necessary to define two-dimensional
wavelets on a sphere. A treatment of the
continuous spherical wavelet transform and
its discretization is given by Freeden \& Winheuser (1996), and an 
algorithm for calculating bi-orthogonal discrete wavelet
bases on a sphere is discussed by Schr\"oder \& Sweldens (1998).
In addition to their possible use in detecting non-Gaussianity
in all-sky CMB maps, the application of such wavelets to the general
analysis of CMB data may also provide significant advantages in the 
compression of large data sets from the forthcoming MAP and Planck Surveyor satellite
missions. These topics will also be investigated in a future paper.

\section*{Acknowledgements}
 
We thank N.~Turok and C.~Barnes for some illuminating preliminary
discussions concerning non-Gaussianity and wavelets. We also thank
J.~Magueijo for useful advice about cumulants, F.R.~Bouchet for
providing the cosmic strings map used in the simulations and
J.~Hawthorn for suggesting the use of Minkowski
functionals to detect non-Gaussianity.
AWJ acknowledges King's College, Cambridge, for financial
support in the form of a Research Fellowship.

\appendix
\section{Moments, cumulants and \lowercase{\textbfit{k}}-statistics}

For a detailed discussion of moments, cumulants and $k$-statistics the
reader is referred to Struart \& Ord (1994) or Kenney \& Keeping
(1954). Here, we outline only the basic definitions.

Let us consider the probability distribution $p(x)$ of some random
variable $x$.  The most straightforward
description of $p(x)$ is in terms of its moments
\[
\nu_r = \langle x^r \rangle,
\]
where the angle brackets denote an ensemble average.
Alternatively, one could describe the distribution using the 
central moments
\[
\mu_r = \langle (x-\nu_1)^r \rangle.
\]
Clearly, if the mean $\nu_1$ of $p(x)$ is zero, then
$\mu_r \equiv \nu_r$. The moments of a distribution are most 
conveniently described the
its moment generating function (MGF), which is given by
\begin{equation}
M(h) \equiv \langle \exp(hx) \rangle = 1+\nu_1h+\nu_2\frac{h^2}{2!}
+\cdots + \nu_r\frac{h^r}{r!} + \cdots.
\label{mgf}
\end{equation}
For example, the MGF of a Gaussian distribution with
mean $\nu_1$ and variance $\sigma^2$ is
\begin{equation}
M(h)=\exp(\nu_1 h +{\textstyle\frac{1}{2}}\sigma^2h^2).
\label{mgfgauss}
\end{equation}

Although the moments (or central moments) 
provide an intuitively straightforward
description of the probability distribution
$p(x)$, an alternative description in terms of
its {\em cumulants} has several advantages. The cumulants $\kappa_r$
of $p(x)$ are defined through the cumulant generating function (CGF) 
\begin{equation}
K(h) \equiv \ln M(h) \equiv \kappa_1 h + \kappa_2\frac{h^2}{2!}+\cdots
+ \kappa_r\frac{h^r}{r!}+\cdots
\label{cgf}
\end{equation}
By comparing (\ref{mgf}) \& (\ref{cgf}), the cumulants
of $p(x)$ are given in terms of its moments by
(Stuart \& Ord 1994)
\begin{equation}
\kappa_r=(-1)^{r-1}
\left|
\begin{array}{ccccc}
\nu_1 & \nu_2 & \nu_3 & \cdots & \nu_r \\
1 & \left({1 \atop 0}\right) \nu_1 & \left({2 \atop 0}\right) \nu_2  &
\cdots & \left({r-1 \atop 0}\right) \nu_{r-1} \\ 
0 & 1 & \left({2 \atop 1}\right) \nu_1 & \cdots &
\left({r-1 \atop 1}\right) \nu_{r-2} \\
0 & 0 & 1 & \cdots & \left({r-1 \atop 2}\right) \nu_{r-3} \\
\vdots & \vdots & \vdots & & \vdots \\
0 & 0 & 0 & \cdots & \left({r-1 \atop r-2}\right) \nu_1
\end{array}
\right|.
\label{determinant}
\end{equation}
Evaluating this determinant for the first four cumulants, we find
\begin{eqnarray*}
\kappa_1 & = & \nu_1 \\
\kappa_2 & = & \nu_2-\nu_1^2 ~=~ \mu_2 \\
\kappa_3 & = & \nu_3-3\nu_2\nu_1+2\nu_1^3 ~=~ \mu_3 \\
\kappa_4 & = & \nu_4-4\nu_3\nu_1+6\nu_2\nu_1^2-3\nu_1^4 ~=~
\mu_4-3\mu_2^2.
\end{eqnarray*}

There are clear advantages to using cumulants rather than moments to
describe a probability distribution. Firstly, for two independent
random variables $x$ and $y$, the cumulants are additive, i.e.
\[
\kappa_r[x+y]=\kappa_r[x] +\kappa_r[y].
\]
Secondly, from
(\ref{mgfgauss}), we see that the CGF of a Gaussian distribution of
mean $\nu_1$ and variance $\sigma^2$ is simply 
\[
K(h)=\nu_1 h+{\textstyle\frac{1}{2}}\sigma^2h^2
\]
Thus for a Gaussian distribution $\kappa_r=0$ for $r>2$. 

From the cumulants and central moments, we can also 
define the dimensionless quantities
\[
\chi_r = \kappa_r/\mu_2^{r/2}
\]
of which $\chi_3$ and $\chi_4$ are better known as the skewness and
kurtosis of the distribution. For a Gaussian distribution, $\chi_3=0$
and $\chi_4=3$, and so it is often useful to define the excess kurtosis
${\cal K}=\chi_4-3$, which equals zero for a Gaussian distribution.

Let us now consider a sample $x_1,x_2,\ldots,x_N$ of size $N$ drawn
from the distribution $p(x)$. Clearly, these
sample values may be used to obtain estimates of the 
cumulants of the parent distribution. The most straightforward method
is first to estimate the moments or central moments of the distribution using
\begin{eqnarray*}
\hat{\nu}_r & = & \frac{1}{N}\sum_{i=1}^N x_i^r, \\
\hat{\mu}_r & = & \frac{1}{N}\sum_{i=1}^N (x_i-\hat{\nu}_1)^r,
\end{eqnarray*}
and then use (\ref{determinant}) to obtain estimates $\hat{\kappa}_r$
of the cumulants.
It is straightforward to show, however, that these estimates are
biassed, so that $\langle\hat{\kappa}_r\rangle \neq \kappa_r$, and this bias
is quite pronounced when the sample size $N$ is small. It is therefore
better to estimate the cumulants of a distribution using
$k$-{\em statistics}. The first four $k$-statistics are given by
\begin{eqnarray*}
\hat{\kappa}_1 & = & \hat{\nu}_1 \\
\hat{\kappa}_2 & = & N\hat{\mu}_2/(N-1) \\
\hat{\kappa}_3 & = & N^2\hat{\mu}_3/(N-1)(N-2) \\
\hat{\kappa}_4 & = & N^2[(N+1)\hat{\mu}_4-3(N-1)\hat{\mu}_2^2]/
(N-1)(N-2)(N-3).
\end{eqnarray*}
The higher-order $k$-statistics are listed by Stuart \& Ord (1994) up
to $r=8$ but are extremely lengthy to write out in full.
Using exact sampling theory, it is simple to show that these new
estimators of the cumulants are unbiassed.

Finally, we note that, if the parent probability distribution $p(x)$ has zero
mean, i.e. $\nu_1=0$ (as is the case for temperature fluctuations in
the CMB), then there exists
an alternative set of unbiassed estimators for the cumulants, which
are much more straightforward to calculate. In this case, the first four
estimators are given by
\begin{eqnarray*}
\hat{\kappa}_1 & = & \hat{\nu}_1 \\
\hat{\kappa}_2 & = & \hat{\nu}_2 \\
\hat{\kappa}_3 & = & \hat{\nu}_3 \\
\hat{\kappa}_4 & = & [(N+2)\hat{\nu}_4-3N\hat{\nu}_2^2]/(N-1).
\end{eqnarray*}

\bsp  
\label{lastpage}
\end{document}